\documentstyle[epsfig]{mn}

\newcommand{\wave}{\langle w\rangle_z}
\newcommand{\wsqr}{\langle w^2\rangle_z}
\newcommand{\zave}{\langle z\rangle}
\newcommand{\epss}{\epsilon_s}
\newcommand{\exep}{\langle\epsilon\rangle_{\epss}}

\title{Constraining the Mass Distribution of Cluster Galaxies by Weak Lensing}
\author[Bernhard Geiger and Peter Schneider]
       {Bernhard Geiger$^{1,2}$ and Peter Schneider$^1$\\
        $^1$Max-Planck-Institut f\"ur Astrophysik,
        Karl-Schwarzschild-Stra{\ss}e 1,
	85740 Garching bei M\"unchen, Germany\\
	$^2$Institut d'Astrophysique de Paris,
	98bis Boulevard Arago, 75014 Paris, France}

\date{Accepted . Received }
\pagerange{\pageref{firstpage}--\pageref{lastpage}}
\pubyear{1997}

\begin{document}

\label{firstpage}

\maketitle

\begin{abstract}
Analysing the weak lensing distortions of the images of faint
background galaxies provides a means to constrain the average mass
distribution of cluster galaxies and potentially to test the extent of
their dark matter haloes as a function of the density of their
environment. The observable image distortions are a consequence of the
interplay between the effects of a global cluster mass distribution
and the perturbations due to individual cluster galaxies.  Starting
from a reconstruction of the cluster mass distribution with 
conventional techniques, we apply a maximum likelihood method to infer
the average properties of an ensemble of cluster galaxies. From
simulations this approach is found to be reliable as long as the
galaxies including their dark matter haloes only contribute a small
fraction to the total mass of the system. If their haloes are
extended, the galaxies contain a substantial mass fraction. In this
case our method is still applicable in the outer regions of clusters,
where the surface mass density is low, but yields biased estimates of
the parameters describing the mass profiles of the cluster galaxies in
the central part of the cluster. In that case it will be 
necessary to resort to more sophisticated strategies by modelling 
cluster galaxies and an underlying global mass distribution 
simultaneously. We conclude that galaxy-galaxy lensing in clusters
provides a unique means to probe the presence and extent of dark haloes
of cluster galaxies.
\end{abstract}


\section{introduction}
Measurements of the rotation curves of spiral galaxies indicate that
they are embedded in massive dark matter haloes. Although less
straightforward, similar kinematical studies for elliptical galaxies
point in the same direction, and are reinforced by a quantitative
statistical analysis of surveys of gravitationally lensed QSOs (Maoz
\&~Rix 1993). The deflection of light rays through the gravitational
action of mass concentrations, usually called gravitational lensing,
provides a way to obtain information about the mass distribution of
galaxies at large radial distances from their centre. [At such
distances, the only luminous test particles to investigate the gravitational 
potential are satellite galaxies with which the haloes of (field) galaxies 
can be probed to $\sim200$kpc radius (Zaritsky \&~White 1994, 
Zaritsky et~al. 1997).] The light deflection causes small distortions of the 
images of faint background galaxies. Recent statistical analyses (Brainerd, 
Blandford \&~Smail 1996, Griffiths et~al. 1996) of these weak distortion 
effects suggest that the dark haloes of (field) galaxies are indeed fairly
extended, as some popular theories of structure formation predict them
to be. During the formation of galaxy clusters the extended haloes of
galaxies may be stripped off due to tidal forces of the cluster
potential or during encounters with other galaxies. Ultimately the
individual galaxy haloes should merge and form a global cluster
halo. In this paper we discuss how this merging picture could be
tested observationally by exploiting weak lensing effects.

The distortions of the images of background galaxies produced by
massive galaxy clusters are strong enough to allow a parameter-free
reconstruction of the clusters' surface mass density. In the last few
years several algorithms of this kind have been developed (e.g.,
Kaiser \&~Squires 1993, Seitz \&~Schneider 1995, 1996, Bartelmann
et~al. 1996, Squires \&~Kaiser 1996) and successfully applied (e.g.,
Fahlman et~al. 1994, Squires et~al. 1996, Seitz et~al. 1996, Fischer
et~al. 1997). The smoothing length which has to be implemented in
these techniques, however, is larger than galaxy scales, and the
amount of information available does not suffice to reconstruct
cluster galaxies individually. Therefore, one has to superpose the
effects of a large number of galaxies statistically in order to infer
the average properties of an ensemble of galaxies. In clusters the
advantages -- compared to `galaxy-galaxy lensing' studies in the field
-- are the larger number density of lens galaxies and the general
amplification of their lensing effects caused by the underlying
cluster mass distribution. However, the analysis is significantly more
difficult technically, because it is necessary to disentangle the
contribution to the image distortions due to individual cluster
galaxies from those of the global cluster mass distribution.

In Section~\ref{simulations} we present simulations of a galaxy cluster, which 
are sufficiently realistic for the purposes of this work, and discuss how 
individual galaxies modify the distortion pattern of a smooth cluster mass 
distribution. In Section~\ref{zeta-sec} we investigate the applicability of 
the so-called $\zeta$-statistic for obtaining information about the cluster 
galaxies. Section~\ref{ml-meth} represents the main part of this paper. Here 
we describe a maximum likelihood method for constraining the mass distribution 
of cluster galaxies, which is based on a reconstruction of the cluster mass 
distribution according to the methods mentioned above. Finally, the results 
are summarized in Section~\ref{discuss}. 

Recently, the weak lensing effects induced by cluster galaxies were
also discussed by Natarajan \&~Kneib (1997). We will comment below on
some of the differences between their approach and ours.

\section{simulations}\label{simulations}

\subsection{Cluster and Cluster Galaxies}
We selected a galaxy cluster located at a redshift $z_{\rm d}=0.16$
from numerical cold-dark-matter simulations (Bartelmann, Steinmetz
\&~Weiss 1995).  In order to produce a map of the surface mass density
$\Sigma$ from the positions of the N-body particles, a smoothing
procedure on scales of about $15\arcsec$ was employed.  We consider a
square field of view with a side length of $10\arcmin$, which
corresponds to a physical size of $1.08h^{-1}\,\rm{Mpc}$ at the
cluster redshift ($\Omega=1$, $\Lambda=0$, and $H_0=100h\,{\rm
km\,s^{-1}Mpc^{-1}}$).  The total mass within that region is
$5.4\times 10^{14}h^{-1}\,{\rm M}_{\sun}$.

For lensing purposes it is convenient to express the surface mass density in
dimensionless form as
\begin{equation}\label{kacrit}
\kappa=\frac{\Sigma}{\Sigma_{\rm crit}}\ \ \ \ \mbox{with}\ \ \ \
\Sigma_{\rm crit}=\frac{c^2}{4\pi\,G}
\frac{D_{\rm s}}{D_{\rm d}\,D_{\rm ds}}\;.
\end{equation}
Here $D_{\rm s}$, $D_{\rm d}$, and $D_{\rm ds}$ are the angular
diameter distances from the observer to the lensed sources, from the
observer to the lensing mass distribution, and between the lens and
the sources, respectively.  Locally, the lensing properties are
specified by the dimensionless surface mass density $\kappa$ and the
shear $\gamma$, which are combinations of second order derivatives of
a common two-dimensional scalar deflection potential. The shear is a
dimensionless two-component quantity, regarded as a complex number
in this paper, and can be computed by integrating over the surface
mass distribution with an appropriate kernel.

In order to populate our cluster with galaxies, the 
following requirements were specified:
\begin{enumerate}
\renewcommand{\theenumi}{(\arabic{enumi})}
\item{The total mass-to-light ratio of the cluster was chosen to be
$300h\,{\rm M}_{\sun}/{\rm L}_{\sun}$.}
\item{Galaxy luminosities $L$ were drawn from a Schechter function 
$\Phi(L)\propto(L/L_{\star})^{\alpha}\,{\rm e}^{-(L/L_{\star})}$ with
$L_{\star}=10^{10}h^{-2}\,{\rm L}_{\sun}$, $\alpha=-1$, and a lower cutoff at 
$0.1\,L_{\star}$.}
\item{Galaxy positions were randomly drawn from those of the N-body
particles, so that `mass follows light' in our model cluster.}
\end{enumerate} 
This procedure resulted in a rich cluster of approximately 360 galaxies, 
about 45 of which are brighter than $L_{\star}$. The exact numbers vary for
different random realizations.

For the mass distribution of the cluster galaxies, a simple truncated 
isothermal sphere model (Brainerd et~al. 1996) was used.
The surface mass density $\kappa$ as a function of the projected radial
distance $\xi$ from the galaxy centre is given by 
\begin{equation}\label{iso}
\kappa(\xi)=\frac{4\pi\,\sigma^2}{c^2}
\frac{D_{\rm d}\,D_{\rm ds}}{D_{\rm s}}\,
\frac{1}{2\xi}\left(1 - \frac{\xi}{\sqrt{s^2+\xi^2}}\right)\;,
\end{equation}
and the cumulative mass can be calculated according to
\begin{equation}
M(<\xi)=\frac{\pi\,\sigma^2}{G}\,\left(\xi+s-\sqrt{s^2+\xi^2}\right)\;.
\end{equation}
The two parameters, velocity dispersion $\sigma$ and cutoff radius $s$,
were chosen as functions of the galaxy luminosity according to the following 
scaling relations:
\begin{equation}\label{scal}
\sigma=\sigma_{\star}\,\left(\frac{L}{L_{\star}}\right)^{1/\eta}\ \ \ \
\mbox{and}\ \ \ \ s=s_{\star}\,\left(\frac{L}{L_{\star}}\right)^{\nu}\;.
\end{equation}
For the first of these relations, which is motivated by the observed
Tully-Fisher and Faber-Jackson relations, a value of $\eta=4$ was used for the 
scaling index. For simplicity, no distinction between spiral and elliptical 
cluster galaxies was made, and the velocity dispersion $\sigma_{\star}$ of an
$L_{\star}$-galaxy was fixed at an intermediate value of $200\,\rm{km/s}$, 
which is closer to the value for early-type galaxies in order to take into 
account their dominance within galaxy clusters. The scaling relation for the 
cutoff radius is more conjectural, and choosing $\nu=0.5$ yields a 
mass-to-light ratio for the galaxies which is independent of
luminosity. Another plausible parametrization would be to assume that the 
cutoff occurs at a fixed density of the dark matter halo, which would give 
$s\propto\sigma$. In order to test the performance of our analysis methods, 
which will be described in the next sections, we specified two models with 
different cutoff radii for the galaxy mass distribution. Choosing a low value 
of $s_{\star}=3.4h^{-1}\,\rm{kpc}$ gives a total $L_{\star}$-galaxy mass of 
$M_{\star}=10^{11}h^{-1}\,\rm{M}_{\sun}$, corresponding to a total 
mass-to-light ratio of $10h$ (in solar units) for the galaxies,
whereas an extended halo of $s_{\star}=34h^{-1}\,\rm{kpc}$ results in
$M_{\star}=10^{12}h^{-1}\,\rm{M}_{\sun}$ and a galaxy mass-to-light ratio
of $100h$. For each of the two cases, the galaxy mass distributions were 
added to the global mass distribution from the numerical cluster simulation, 
which had been scaled such that the total mass of the system remains constant. 
Fig.\,\ref{cluster} displays contour plots of the resulting surface mass 
density according to the two different galaxy models and illustrates their 
differences regarding the structure of the dark matter distribution. 
\begin{figure*}
    \vspace{150mm}
%    \epsfxsize=150mm
%    \epsffile{cluster.eps} 
\caption{The mass distribution for the simulated cluster of galaxies.
The {\bf top} panels show contours of the surface mass density corresponding 
to the two different models for the galaxy mass distribution. The 
{\bf left} plot is for a cutoff radius of $s_{\star}=3.4h^{-1}\,\rm{kpc}$, 
and the {\bf right} one for $s_{\star}=34h^{-1}\,\rm{kpc}$. Respectively, the 
{\bf bottom} panels show the pattern of the average ellipticities of background
galaxy images, overlaid with the reconstructed cluster mass distribution. 
The largest marks in these plots, indicating the strongest distortion effects,
represent an average ellipticity of $|\overline\epsilon|\approx0.4$.
For clarity, the ellipticity patterns are displayed on a $20\times 20$ grid, 
whereas the reconstructions were calculated from a $30\times 30$ grid. Note 
that the same realization of background galaxies with the same intrinsic
ellipticities was used for both cases, and therefore the reconstructed mass 
maps contain similar noise properties. The field of view is $10\farcm05$ and 
the contours are $\kappa_{\infty}=0.05,0.1,0.2,0.3,0.4,0.5,0.6,\dots,1.2$.}
\label{cluster}
\end{figure*}

Similarly, the shear corresponding to the total mass distribution can be
obtained by adding the shear contribution from the individual galaxies
to the (scaled) shear map 
calculated from the original surface mass distribution of the cluster. For the 
galaxy mass model of equation~(\ref{iso}), the modulus of the shear is given by
\begin{equation}\label{gal-shear}
|\gamma|(\xi)=\frac{4\pi\,\sigma^2}{c^2}
\frac{D_{\rm d}\,D_{\rm ds}}{D_{\rm s}}\,\frac{1}{2\,\xi}\,
\left(1+\frac{2\,s}{\xi}-\frac{2\,s^2+\xi^2}{\xi\sqrt{s^2+\xi^2}}\right)\;,
\end{equation}
and its phase can be determined from the position angle with respect to the 
galaxy centre.

\subsection{Distortion Effects and Background Galaxies}
For describing the ellipticities of galaxy images we use the complex parameter
$\epsilon$, which is defined in terms of the second moments of the image
brightness distribution (see, e.g., Seitz \&~Schneider 1997). In the special 
case of elliptical isophotes with axis ratio $r\leq1$, its modulus is given by 
$|\epsilon|=(1-r)/(1+r)$. A graphical visualization of the ellipticity
parameter space can be found in Fig.\,\ref{pddeps}. The distortion
effects exerted by the lens on the  
images of background galaxies do not depend on the parameters $\kappa$ and 
$\gamma$ individually, but only on the combined quantity `reduced shear' 
$g=\gamma/(1-\kappa)$. The transformation of an intrinsic galaxy ellipticity 
$\epss$ to the observable image ellipticity $\epsilon$ is given by 
\begin{equation}\label{epssima}
\epsilon(\epss\,|\,g)=
\left\{
\begin{array}{lll}
\frac{\displaystyle \epss+g}
{\displaystyle 1+g^{\star}\epss} & {\rm for} & |g|\leq 1 \\
\frac{\displaystyle 1+g\,\epss^{\star}}
{\displaystyle \epss^{\star}+g^{\star}} & {\rm for} & |g|>1\;,
\end{array}
\right.
\end{equation}
with a case distinction between the even-parity ($|g|<1$) and the 
odd-parity ($|g|>1$) regions. 

Fig.\,\ref{distortion} shows a map of the modulus of the reduced shear
for the simulated galaxy cluster described above. In the even-parity
region this quantity is a measure for the strength of the distortion
effects. The global mass distribution is non-critical, which means
that $|g|<1$ everywhere, except very close to the centre of the individual 
cluster galaxies. In general, the image distortions
tend to be aligned tangentially towards the centre of mass
concentrations. The figure illustrates the perturbing effects of the
individual cluster galaxies. At their positions in a radial direction
towards and away from the cluster centre, the strength of the
distortion is locally increased because the effects of the global
cluster mass distribution and the individual galaxies then act in the
same direction. But in the direction tangential to the cluster centre,
the orientation of the galaxy contribution to the shear is
perpendicular to the shear direction of the cluster, and therefore
these effects partly cancel, leading to a reduction in the strength of
the distortion effects.
\begin{figure*}
    \vspace{150mm}
%    \epsfxsize=150mm
%    \epsffile{distortion.eps} 
\caption{The modulus $|g|$ of the reduced shear for the simulated cluster of
galaxies (with $s_{\star}=34h^{-1}\,\rm{kpc}$). This quantity is a measure 
for the strength of the distortion effects (modulo the restrictions mentioned 
in the text). In order to emphasize the perturbations induced by the 
individual galaxies by increasing the contrast, a saturation value of 
$|g|=0.3$ was used for the grey scales. The plot was calculated for a 
source redshift of $z=1$.}
\label{distortion}
\end{figure*}

However, Nature does not provide us with a continuous map of the lensing
properties, but only with very noisy estimates of the reduced shear at the 
discrete positions of background galaxy images. For these simulations, a 
population of background galaxies was generated with a number density 
of $40/\rm{arcmin}^2$. Their intrinsic ellipticities were drawn from a 
probability distribution of the form
\begin{equation}\label{pddint}
p_{\epss}(\epss)=
\frac{1}{\pi\,\sigma_{\epss}^2\,(1-{\rm e}^{-1/\sigma_{\epss}^2})}\,
{\rm e}^{-(|\epss|/\sigma_{\epss})^2}
\end{equation}
with dispersion $\sigma_{\epss}=0.2$, and their positions were randomly
distributed within the field of view.

Up to now we did not specify the redshifts of the source galaxies, and
the discussion above assumed them to be located in a single redshift plane. 
The strength of the lensing effect depends on the source redshift via the 
angular diameter distances appearing in the definition of the critical surface 
mass density in equation~(\ref{kacrit}). We adopt the approach of 
Seitz \&~Schneider (1997) and relate the lensing parameters to their 
respective values corresponding to (hypothetical) sources located at infinite 
redshift. The surface mass density and the shear as a function of source 
redshift $z$ can then be expressed as $\kappa(z)=w(z)\,\kappa_{\infty}$ and 
$\gamma(z)=w(z)\,\gamma_{\infty}$, and for an Einstein-de~Sitter universe the 
`relative lensing strength' can be calculated as 
\begin{equation}\label{lensstr}
w(z)=
\left\{
\begin{array}{cll}
0 & {\rm for} & z\leq z_{\rm d} \\
\frac{\displaystyle\sqrt{1+z}-\sqrt{1+z_{\rm d}}}
{\displaystyle\sqrt{1+z}-1}  & {\rm for} & z>z_{\rm d}\;.
\end{array}
\right.
\end{equation}
The reduced shear as a function of source redshift is given by
\begin{equation}\label{gvonz}
g(z\,|\,\kappa_{\infty},\,\gamma_{\infty})=
\frac{w(z)\,\gamma_{\infty}}{1-w(z)\,\kappa_{\infty}}\;,
\end{equation}
and therefore the statistical properties of the lensing effect depend on the 
surface mass density $\kappa_{\infty}$ and the shear $\gamma_{\infty}$
explicitly, although the degeneracy mentioned earlier is only weakly broken as
long as the lens is non-critical for all redshifts.
For each of the source galaxies a redshift was drawn from the probability 
distribution (Brainerd et~al. 1996)
\begin{equation}\label{reddist}
p_z(z\,|\,z_0,\,\beta)=\frac{\beta\,z^2}{\Gamma(\frac{3}{\beta})\,z_0^3}\,
{\rm e}^{-\left(z/z_0\right)^\beta}
\end{equation}
with $z_0=1/3$ and $\beta=1$ which results in an average redshift of 
$\zave=1$. The observable image ellipticities were then calculated
by applying equations (\ref{lensstr}), (\ref{gvonz}), and~(\ref{epssima}).
Some of our source galaxies are in fact unlensed foreground objects, but for 
simplicity the entire population generated in this way is referred to as 
`background galaxies'. The lensing effects of background galaxies on the
images of more distant galaxies within that population are neglected in this
study. 

Whenever we use the quantities $\kappa$, $\gamma$, or $g$ without
redshift-argument or $\infty$-subscript in the rest of the paper, this should 
be regarded as referring to a single redshift plane.

\subsection{Cluster Mass Reconstruction}\label{reconstruction}
Assuming that we can unambiguously distinguish the cluster galaxies from the
population of background galaxies, we calculated the average image ellipticity 
$\overline\epsilon$ for the latter on a $30\times 30$ grid using a 
Gaussian smoothing procedure with variable smoothing length in order to 
account for the varying strength of the distortion effects. For simplicity, 
this smoothing length was adjusted linearly from $0\farcm2$ at the cluster 
centre to $1\farcm0$ at the boundary of the field of view, although in 
principle objective strategies could be developed for its optimal choice.   
Fig.\,\ref{cluster} displays the gridded distortion pattern determined in this
way, as well as the reconstruction of the cluster surface mass density 
calculated from it by applying the non-linear finite-field
inversion method described in Seitz \&~Schneider (1996), and taking into 
account the redshift distribution of the sources as explained in Seitz 
\&~Schneider (1997). Here we assumed the true redshift 
distribution to be known, and we will comment on the consequences of giving
up this assumption in Section~\ref{problems}. 

The figure reveals that there are only minor differences in the reconstructed 
cluster mass distribution for the two different galaxy mass models. 
The information contained in the background galaxy images about the structure
of the mass distribution on galaxy scales is very efficiently erased by the 
averaging procedure described above. In reconstructions performed with a 
refined grid and reduced smoothing length for the galaxy input model 
with large cutoff radius, it is sometimes possible to identify reconstructed 
mass clumps with groups of massive galaxies. However, a quantitative analysis
of the significance of sub-clumps in mass maps is extremely difficult, and
therefore this does not seem to be a practical method to constrain the mass
distribution of cluster galaxies.

Analysing the weak distortions of background galaxies only allows
the mass distribution of the lens to be determined 
up to a global invariance transformation of the form 
\begin{equation}\label{mass-sheet}
\kappa(\vec\xi)\rightarrow(1-\kappa_{\rm s})\,\kappa(\vec\xi)
+\kappa_{\rm s}\;,
\end{equation}
which corresponds to adding a sheet of constant surface mass density
$\kappa_{\rm s}$ after appropriately rescaling the reconstructed solution 
(Falco, Gorenstein \&~Shapiro 1985, Schneider \&~Seitz 1995). 
For practical purposes this degeneracy remains even if the sources are
distributed in redshift. In this case transformations similar to
equation~(\ref{mass-sheet}) apply (Seitz \&~Schneider 1997). In our 
simulations we artificially adjusted the reconstructed surface mass density 
such that the total mass of the cluster is correctly reproduced. In practice, 
the `mass sheet degeneracy' can be broken by postulating that 
the surface mass density of a reasonable galaxy cluster should have dropped to 
insignificant values at the boundaries of a large field of view, or by 
exploiting magnification effects, either through
the lensing effects on the number counts of background galaxies 
(Broadhurst, Taylor \&~Peacock 1995) or the size-magnitude relation
(Bartelmann \&~Narayan 1995), which are not invariant under the 
transformation~(\ref{mass-sheet}).

Another technical difficulty is the following: In addition to the reconstructed
surface mass density, the likelihood method to be described in 
Section~\ref{ml-meth} also requires a map of the shear  which corresponds to 
this mass distribution. However, calculating the shear from the surface mass
density {\it a posteriori}\/ involves an integration extending beyond the 
limited data region. Again, there will be no practical problems, if 
the surface mass density attains negligible values at the boundary of the 
field of view, provided that there are no huge mass clumps lurking just 
outside of it.

\section{$\zeta$-statistic}\label{zeta-sec}

\subsection{Theory}
Kaiser (1995) showed that the difference of the average surface mass densities
within a circular aperture $\overline{\kappa}(x_1)$ and an annulus around 
that aperture $\overline{\kappa}(x_1,x_2)$ can be calculated from the shear
within the annulus:
\begin{equation}\label{zeta-def}
\zeta(x_1,x_2):=
\overline{\kappa}(x_1)-\overline{\kappa}(x_1,x_2)=
\frac{2\,x_2^2}{x_2^2-x_1^2}\int\limits_{x_1}^{x_2}\frac{{\rm d}x}{x}\,
\langle\gamma_{\rm t}\rangle(x)\;.
\end{equation}
The variable $x$ represents a radial coordinate measured from the centre of the
aperture, and $x_1$ and $x_2$ denote the inner and the outer radius of the
annulus, respectively. $\gamma_{\rm t}$ is the tangential component of the
shear and $\langle\gamma_{\rm t}\rangle(x)$ is its circularly averaged value
as a function of the radial distance. This equation was first applied by
Fahlman et~al. (1994) in order to determine a rigorous lower limit on the
mass of the galaxy cluster MS1224, without the need to worry about
non-linear lensing 
properties or the confusion of background and cluster galaxies in the cluster
centre. In this section we investigate the possibilities offered by this method
for obtaining information on the mass distribution of cluster
galaxies. This can be achieved by analysing the distortion of background 
galaxy images in annuli centred on individual cluster galaxies and adding the 
effects of a large number of them in order to get a significant signal.  
A nice feature of this application of relation (\ref{zeta-def}) is that the
reference to the surface mass density in the annulus automatically takes into
account an underlying cluster mass distribution and directly
measures the galaxy masses, provided the surface mass density of the
cluster can be reasonably approximated locally as a linear
function. It is easy to see that a linear 
trend in the cluster mass profile does not affect $\overline{\kappa}(x_1,x_2)$
and so only higher order variations of the cluster mass distribution on scales
comparable to the size of the annulus could bias the mass measurement.

The right-hand-side of equation~(\ref{zeta-def}) can be written as a
two-dimensional integral and, therefore, be approximated by a sum over the
discrete data points which are provided by the images of background galaxies: 
\begin{equation}\label{zeta-sum}
\zeta(x_1,x_2)
\approx x_2^2\;\frac{1}{N}\,\sum_{i=1}^{N}\,\frac{\gamma_{{\rm t}i}}{x_i^2}
\approx x_2^2\;\frac{1}{N}\,\sum_{i=1}^{N}\,
\frac{\epsilon_{{\rm t}i}\,(1-\kappa_i)}{x_i^2}\;.
\end{equation}
With our definition for the ellipticity parameter $\epsilon$, the expectation
value for observed image ellipticities is equal to the reduced shear: 
$\exep=g$ (Schramm \&~Kayser 1995, Seitz \&~Schneider 1997). Therefore, 
each observed image ellipticity $\epsilon_i$ is an unbiased -- though very
noisy -- estimate for the reduced shear $g_i=\gamma_i/(1-\kappa_i)$ at the 
image position, and $\gamma_{{\rm t}i}$ can be replaced by 
$\epsilon_{{\rm t}i}\,(1-\kappa_i)$ in the above equation. (Here we 
restricted the treatment to the even-parity region; in the odd-parity case 
$\exep=1/g^{\ast}$.) In the limit $\kappa\ll 1$ the shear can be directly 
estimated from the image ellipticities ($\exep\approx\gamma$) 
and no further information about the cluster mass distribution is required 
for applying the $\zeta$-statistic. When leaving the linear regime, however, 
the corrective factor ($1-\kappa$) becomes important. In this case, the 
surface mass density $\kappa_i$ at the image positions can be taken from a 
reconstruction of the cluster mass distribution.\footnote{The `radial averaging
method' employed by Natarajan and Kneib (1997) is similar to the one described
here, but it does not include the extension into the non-linear lensing regime
provided by the $(1-\kappa)$-factor. Whereas this omission does not preclude
the significant detection of a lensing signal by cluster galaxies, it renders a
quantitative interpretation more indirect.} [Performing a mass sheet 
transformation of the reconstructed mass distribution according to 
equation~(\ref{mass-sheet}) modifies $\zeta$ and all galaxy mass estimates 
derived from it by a factor $(1-\kappa_{\rm s})$. This can easily be seen by 
replacing $\kappa_i$ with $(1-\kappa_{\rm s})\,\kappa_i+\kappa_{\rm s}$ in 
equation~(\ref{zeta-sum}).]

The calculation of $\zeta$ according to equation~(\ref{zeta-sum}) can be
regarded as a kind of noisy Monte-Carlo integration. Both of the two
approximate-equality signs only hold for a rather large number $N$ of 
background images and become equalities for $N\rightarrow\infty$. They 
express two different kinds of uncertainties; the first one those which 
are arising from sparse sampling of the integration area, and the second 
one those from the noisy data points. The errors in $\zeta$ due to the latter
may be expressed in terms of the intrinsic ellipticity dispersion
$\sigma_{\epss}$ as 
\begin{equation}\label{zeta-sig}
\sigma[\zeta(x_1,x_2)]\approx x_2^2\;\frac{\sigma_{\epss}}{\sqrt{2}}
\;\frac{1}{N}\,\left(\sum_{i=1}^{N}\,\frac{(1-\kappa_i)^2}{x_i^4}\right)
^\frac{1}{2}\;,
\end{equation}
which is not quite exact, because lensing changes the dispersion of the  
probability distribution for the observed image ellipticities (see 
Section~\ref{probability}). Due to the rather inhomogeneous shear pattern 
which is induced by the cluster galaxies (see Fig.\,\ref{distortion}), the 
errors from the sparse sampling of the integration area may also be 
substantial in our application of the $\zeta$-statistic, and as $\kappa_i$ 
approaches unity those introduced by an inaccurate reconstruction of the 
cluster mass distribution will become important as well.
	
The discussion so far assumed a single redshift plane for the background
sources. In the case of a redshift distribution, the equations are still valid 
for $\kappa_i\ll 1$, if the surface mass density and the shear are interpreted 
as quantities referring to the redshift-averaged critical surface mass
density, which means $\kappa=\wave\,\kappa_{\infty}$ and 
$\gamma=\wave\,\gamma_{\infty}$. But in the non-linear regime,
the treatment should be generalized, because then the expectation value for
observed image ellipticities also depends on higher moments of the 
redshift distribution. Seitz \&~Schneider (1997) derived the approximation
\begin{equation}\label{eps-appr}
\langle\epsilon\rangle_{\epss,\,z}\approx
\frac{\wave\,\gamma_{\infty}}{1-\frac{\wsqr}{\wave}\,\kappa_{\infty}}\;,
\end{equation}
which is quite accurate for $\kappa_{\infty}\la 0.8$ and generic redshift
distributions. In analogy to the calculations above, this offers an estimate 
for the tangential shear in terms of the observed image ellipticities (and the
reconstructed cluster mass distribution), and expressions equivalent to 
equations (\ref{zeta-sum}) and~(\ref{zeta-sig}) can be computed. In the 
application to the simulations we employed this slightly generalized 
formalism. 

\subsection{Application to Simulations}
The radial mass profile of the cluster galaxies can be probed by 
calculating the $\zeta$-statistic as a function of the inner radius of the 
annulus. The minimal inner radius for applying this method is limited by the 
ability to measure reliable ellipticities for background galaxy images in 
the vicinity of the typically much brighter cluster galaxies. Another, 
theoretical complication for images located very close to cluster galaxies 
is that those also contribute to the surface mass density at the image 
positions, and this should in principle be included in the corrective 
factor ($1-\kappa_i$) as well, which requires specifying a model for the galaxy
mass distribution. However, this problem can be neglected in view of the 
observational limitations mentioned above, because at useful radial distances 
the surface mass density of the cluster galaxies should already have dropped 
to insignificant values. 

\begin{figure}
    \epsfxsize=84mm
    \epsffile{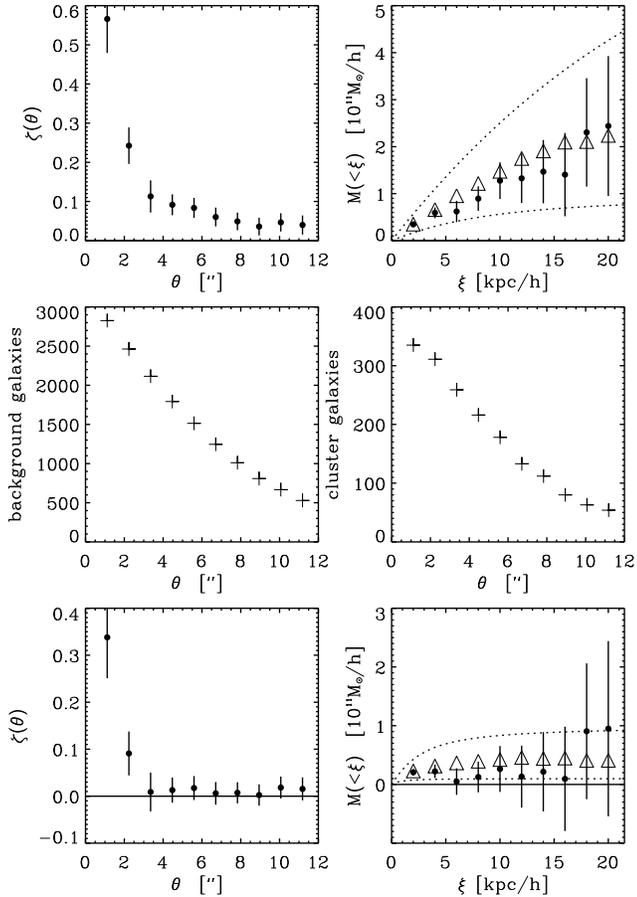}
\caption{Application of the $\zeta$-statistic to the simulations. 
The plots show $\zeta$ and the aperture mass measurements derived from it as a 
function of the (inner) aperture radius. 
The {\bf top} panels are for the input galaxy mass model with a large 
cutoff radius of $s_{\star}=34h^{-1}\,{\rm kpc}$ and the {\bf bottom} panels 
for the model without extended dark matter halo 
($s_{\star}=3.4h^{-1}\,{\rm kpc}$). Regardless of their luminosity, cluster 
galaxies were included in the analysis according to the criterion described 
in the text. The triangles denote the average mass of the included cluster 
galaxies and therefore represent the quantity which is supposed to be measured 
by the lensing estimates. For orientation, the dotted lines show the mass 
profile of an $0.1\,L_{\star}$- and an $L_{\star}$-galaxy according to the 
respective input model. The {\bf middle} panels display the number of 
background and cluster galaxies used in the analysis.}
\label{zeta1}
\end{figure}
\begin{figure}
    \epsfxsize=84mm
    \epsffile{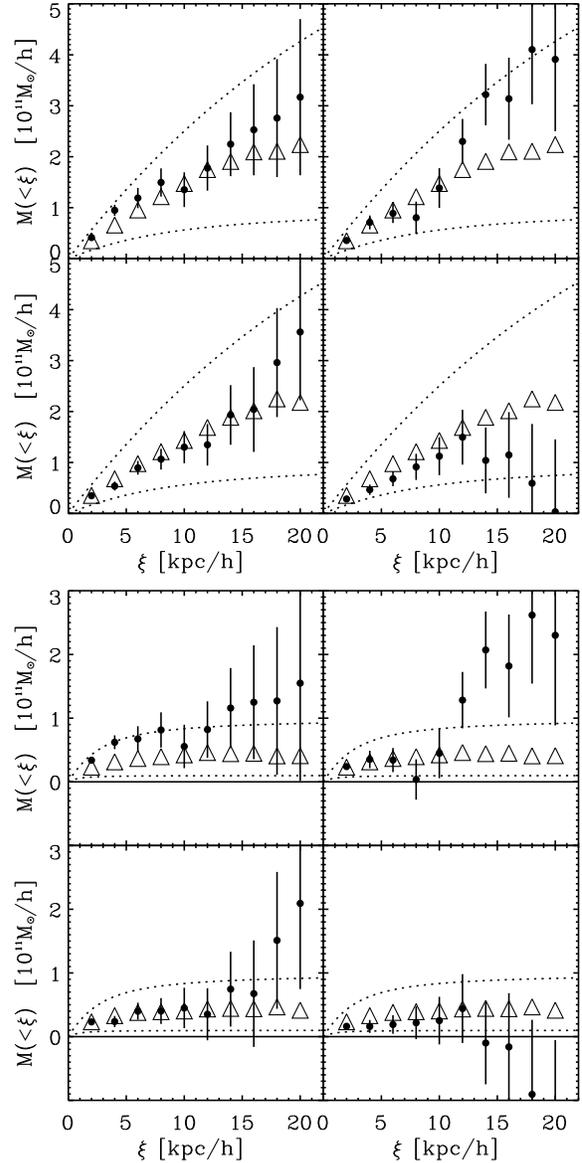}
\caption{Mass estimates from the $\zeta$-statistic for different realizations 
of cluster and background galaxies. The four diagrams at the {\bf top} are 
for the input model with large cutoff radius and the {\bf bottom} ones for the 
small cutoff radius. (As mentioned in the text, the innermost data point in 
these plots can be systematically affected by a non-negligible cluster galaxy 
contribution to the surface mass density at the image positions, and in 
practice, the determination of this data point will be hampered by 
observational problems.)}
\label{zeta2}
\end{figure}
The outer radius of the annulus must not be so large that the reference-term 
$\overline{\kappa}(x_1,x_2)$ picks up variations of the cluster mass 
distribution. In practice, a more serious limitation for its extent 
is the presence of neighbouring cluster galaxies, which must not be located 
within the annulus in order to keep $\overline{\kappa}(x_1,x_2)$ as
small as possible. As long as these constraints are satisfied, 
$\overline{\kappa}(x_1,x_2)$ and therefore also $\zeta(x_1,x_2)$ are nearly
independent of the outer radius $x_2$. This means that the outer radius can 
be chosen individually for each cluster galaxy and still the $\zeta$-estimates 
from each of those can be combined afterwards to achieve significant 
results. In order to use the available information effectively, we therefore
adopted the following strategy for calculating the mass estimates for a 
given value of the inner radius. Cluster galaxies were included 
in the analysis, if inner-radius circles centred on them did not intersect 
the inner-radius circle drawn around any other cluster galaxy. (This leads to 
a bias of the positions of the included cluster galaxies away from the cluster
centre.) For each of the cluster galaxies used, the outer radius was then 
specified as the maximal radius possible without intersecting the 
{\it inner}\/-radius circle around any other cluster galaxy. With this 
prescription, some background galaxy images located between cluster galaxies 
have to be included in the analysis twice or several times, with reference to 
different cluster galaxies. Note that the method implicitly takes into account 
the shear effects of more than one cluster galaxy on individual background 
images. 

The radial coordinate $x$ used so far can either be regarded as an angular 
separation $\theta$ on the sky or as the projected physical separation 
$\xi=D_{\rm d}\,\theta$ in the lens plane. For a given inner radius, 
the $\zeta$-value can be trivially converted into an estimate of the 
projected mass within that aperture: 
\begin{equation}\label{zeta-mass}
M(<\xi)=\pi\,\xi^2\;\zeta(\xi)\;\Sigma_{\rm crit}\;.
\end{equation}
Fig.\,\ref{zeta1} displays the results of applying the method to the 
simulations described in Section~\ref{simulations}. The plots of this figure 
reveal that the signal-to-noise ratio of the aperture mass measurement 
rapidly deteriorates for increasing aperture radius, because the number of 
background and cluster galaxies which can be used for the analysis then 
considerably decreases. The error bars drawn in the figure, which are of 
course correlated for different data points, were calculated according to 
equation~(\ref{zeta-sig}) and only include the uncertainties due to the 
intrinsic ellipticity distribution of the sources. More comprehensive error 
estimates could be computed from the simulations. In order to give an
indication of the true errors for the mass determination, Fig.\,\ref{zeta2}
depicts the results for different realizations of cluster and background 
galaxies. To be specific, we selected two realizations from a sample 
of five random sets of background galaxies and combined them with two 
different random realizations of cluster galaxies. For the galaxy mass model 
with a large cutoff radius, significant mass detections are feasible up to 
radial distances of $\approx 15h^{-1}\,{\rm kpc}$. For the model with a 
small cutoff radius, however, it is hardly possible to achieve significant 
results at all. Nevertheless, the data still allow us to set limits on the 
presence of an extended dark matter halo in this case.

In the application of the $\zeta$-statistic to the simulations we
mainly concentrated on exploring the capabilities of the method for reliably
retrieving the input values, and in view of the more powerful techniques to be
described in the next section, we do not further discuss the optimal 
strategies for quantifying or interpreting the results provided by it.

The $\zeta$-statistic allows us to determine a direct galaxy mass estimate 
without any model assumptions, and it can be conveniently applied in the 
outskirts of a cluster where the separations between the cluster galaxies 
are large. However, due to geometrical constraints the method cannot 
make optimal use of all the information available; in particular, it is not 
well suited to test the radial extent of the galaxy mass distribution. 
Towards the cluster centre, the crowding of cluster galaxies seriously 
compromises the applicability of the method. In addition to that, 
the generalization to the non-linear regime removes much of the 
$\zeta$-statistic's original simplicity, and an accurate description of the 
global cluster mass distribution becomes important.

\section{maximum likelihood method}\label{ml-meth}

\subsection{Model Specification}\label{mod-spec}
The maximum likelihood method described here follows in part the prescription
of Schneider \&~Rix (1997) for weak lensing by field galaxies. The principle
philosophy of likelihood techniques is to specify a model, calculate the
probability distribution of observable quantities according to the model, and
maximize the joint probability density for the actually observed values by
varying the model parameters. In our case, the observable image distortions 
are a consequence of the interplay between the effects of a global cluster 
potential and the perturbations due to individual galaxies. In addition to 
specifying a parametrized mass model for the galaxies, it is therefore 
important to have an accurate description of the cluster mass distribution 
which is provided by the reconstruction presented in 
Section~\ref{reconstruction}. 

As a model for the galaxy mass distribution we again use the truncated 
isothermal sphere (\ref{iso}). Of course, this model is appropriate for the 
synthetic data used here, whereas one could argue that realistic galaxy haloes 
in clusters might rather be flattened or completely irregular. However, this 
analysis is aimed at determining the average properties of an ensemble of 
galaxies which might still be reasonably well 
described by a simple model with a 
characteristic scale and normalization as parameters. In order to add the 
information from galaxies with different luminosities, the scaling relations
(\ref{scal}) were applied. Adding the mass models for each of the cluster 
galaxies to the cluster reconstruction then yields a model for the total mass 
distribution of the system as a function of the model parameters, which are 
the velocity dispersion $\sigma_{\star}$ and the cutoff radius $s_{\star}$ of 
an $L_{\star}$-galaxy, and the scaling indices $\eta$ and $\nu$. 
Analogously, the shear contribution due to the galaxies, which is given by 
equation~(\ref{gal-shear}), can be added to the global shear from the 
reconstruction (see Section~\ref{reconstruction}) in order to obtain the total 
shear corresponding to the total mass model. 

A complication which has to be taken into account when performing this 
procedure is the following: If the individual galaxies do
have extended haloes, the mass in galaxies constitutes a significant fraction 
of the total cluster mass. The cluster reconstruction is sensitive to 
the total mass, and therefore it already includes the (smoothed-out)
mass contribution from the galaxies. 
This means that the additional mass added by the galaxy models has to be 
compensated in some way. This was done by simply scaling down the 
reconstruction appropriately or by subtracting surplus mass locally at the 
position of cluster galaxies. The merits and limitations of these 
({\it ad hoc}\/) procedures will become evident in Section~\ref{vel-cut}.
   
The total mass model constructed in this way determines the values for the
lensing quantities $\kappa_{\infty}$ and $\gamma_{\infty}$ at the position of 
each background galaxy image as a function of the galaxy model parameters. 
Given the surface mass density and the shear, the next section deals with 
the problem of calculating the probability density distributions for image 
ellipticities.

\subsection{Probability Density Distributions}\label{probability}
In the case of a single redshift plane for the background galaxies, the
probability density for observing an image ellipticity $\epsilon$
is given by 
\begin{equation}\label{pddzdelta}
p_{\epsilon}(\epsilon\,|\,g)=
p_{\epss}(\epss(\epsilon\,|\,g))\,
\left|\frac{{\rm d}^2\epss}{{\rm d}^2\epsilon}\right|
(\epsilon\,|\,g)\;,
\end{equation}
and it is completely specified by the reduced shear at the image 
position.\footnote{Note that the Jacobian determinant in (\ref{pddzdelta})
has been left out in Natarajan \&~Kneib (1997).}
The transformation $\epss(\epsilon\,|\,g)$ of image to source ellipticities 
can be obtained by inverting equation~(\ref{epssima}), and the Jacobian 
determinant of this transformation can be computed according to
\begin{equation}
\left|\frac{{\rm d}^2\epss}{{\rm d}^2\epsilon}\right|(\epsilon\,|\,g)=
\left\{
\begin{array}{lll}
\frac{\displaystyle (|g|^2-1)^2}
{\displaystyle |\epsilon\,g^{\star}-1|^4} & {\rm for} & |g|\leq 1 \\
\frac{\displaystyle (|g|^2-1)^2}
{\displaystyle |\epsilon-g|^4} & {\rm for} & |g|>1\;.
\end{array}
\right.
\end{equation}
[In the limit $|g|\ll1$, the ellipticity transformation reads 
$\epsilon\approx\epss+g$, and the Jacobian determinant 
reduces to unity. The lensed probability density distribution is then
approximately equal to the intrinsic one shifted in the shear direction: 
$p_{\epsilon}(\epsilon\,|\,g)\approx p_{\epss}(\epsilon-g)$.]

However, if the sources are distributed in redshift, the reduced shear has 
to be calculated from $\kappa_{\infty}$ and $\gamma_{\infty}$ as a function of
redshift according to equation~(\ref{gvonz}). In this case, the probability 
density for observing an ellipticity $\epsilon$ can be obtained by integrating 
equation~(\ref{pddzdelta}) over redshift
\begin{eqnarray}\label{pdddist}
\lefteqn{p_{\epsilon}(\epsilon\,|\,\kappa_{\infty},\,\gamma_{\infty})=} \\
& {\displaystyle \int\limits_0^{_{\infty}}{\rm d}z\,p_z(z)\,
p_{\epss}(\epss(\epsilon\,|\,\kappa_{\infty},\,\gamma_{\infty},\,z))
\,\left|\frac{{\rm d}^2\epss}{{\rm d}^2\epsilon}\right|
(\epsilon\,|\,\kappa_{\infty},\,\gamma_{\infty},\,z)} \nonumber 
\end{eqnarray}
and it explicitly depends both on the surface mass density and on the shear.
In order to calculate this probability density in practice, it is necessary 
to know the intrinsic ellipticity distribution $p_{\epss}(\epss)$ of the 
sources, which we assume to be available from observations in `empty fields'. 
(A possible dependence of the intrinsic distribution on the redshift, the 
magnitude, or other galaxy characteristics could easily be taken into account 
in the above equation.) A more serious problem is to specify an estimate for 
the redshift distribution of the sources. As was mentioned in 
Section~\ref{reconstruction} we use the true $p_z(z)$ within our analysis, and 
discuss the consequences of an incorrect redshift distribution in 
Section~\ref{problems}. 

\begin{figure}
    \epsfxsize=84mm
    \epsffile{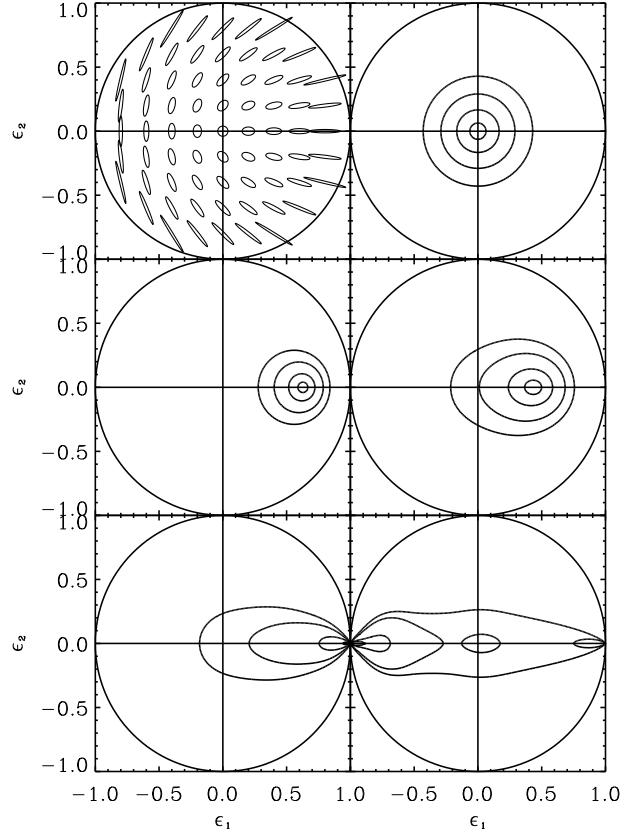}
\caption{The probability density distribution for image ellipticities. 
The {\bf top left} plot is a graphical visualization of the ellipticity 
parameter $\epsilon=\epsilon_1+{\rm i}\,\epsilon_2$. It displays the shape of 
the images of an intrinsically circular source. The {\bf top right} plot
shows the isotropic probability density distribution for the intrinsic shape 
of background galaxies [equation~(\ref{pddint}) with $\sigma_{\epss}=0.2$]. 
The contour lines in this and in the following plots enclose 99\%, 90\%, 50\%, 
and 10\%, respectively, of the probability. The other four diagrams display
the probability density distribution for the image ellipticity of lensed 
background galaxies. The {\bf middle left} plot is for $g=0.6$ and single
redshift sources, and the {\bf middle right} one for $\gamma_{\infty}=0.3$, 
$\kappa_{\infty}=0.5$ and a redshift distribution of the sources according to
equation~(\ref{reddist}) with $\beta=1$ and $\zave=1$. The lens 
redshift is $z_{\rm d}=0.16$. Note that we chose real $g$ and 
$\gamma_{\infty}$ values for this illustration, which specifies the 
$\epsilon_1$-axis as the `shear direction'. The {\bf bottom} plots show the 
ellipticity distributions for lens parameters which are 
`tangentially critical' ({\bf left}, $\gamma_{\infty}=0.4$, 
$\kappa_{\infty}=0.8$) and `radially critical' ({\bf right}, 
$\gamma_{\infty}=0.6$, $\kappa_{\infty}=2.0$) with the same redshift 
distribution as above. (In the bottom plots the 10\%-contour line is not 
visible in this representation.)} 
\label{pddeps}
\end{figure}
Fig.\,\ref{pddeps} illustrates the modification of the ellipticity distribution
induced by lensing. In the case of a fixed source redshift, the lensed 
distribution has the nice property that the expectation value of the image 
ellipticities recovers the reduced shear, which has already been used in 
Section~\ref{zeta-sec}. The dispersion of the ellipticity distribution is 
reduced compared to the intrinsic one, and although the distribution becomes 
skewed, its contour lines remain fairly circular. In fact, it can be shown 
that the dispersion of the distribution in `shear direction' (along the 
$\epsilon_1$-axis in the plot) is equal to the dispersion perpendicular to 
this direction in the ellipticity coordinates (along the $\epsilon_2$-axis in 
the plot). In the presence of a redshift distribution, the probability density
distribution for the image ellipticities becomes elongated along the `shear 
direction', because the strength of the distortion effect depends on the 
redshift of the sources. The images of galaxies located just behind the lens 
are only slightly affected, and obviously foreground objects are not distorted 
at all. 

For completeness, the figure also illustrates the ellipticity 
distribution for lens parameters which are critical for sufficiently high 
redshifts. In this case, $p_{\epss}(\epss)$ formally includes a 
$\delta$-`function' contribution at ellipticity coordinates corresponding to 
arcs. For a `radially critical' lens, images can either be distorted 
tangentially or radially, depending on the source redshift, and in principle a 
radial and a tangential arc could be superposed at the same position on the 
sky.

\subsection{Likelihood Function and Confidence Contours}\label{like-conf}
The likelihood function ${\cal L}$ can now be defined as the product of the 
probability densities of the actually measured ellipticities $\epsilon_i$ of 
all the background galaxy images
\begin{equation}
{\cal L}(\sigma_{\star},\,s_{\star},\,\eta\,,\nu):=
\prod_i\,p_{\epsilon}(\epsilon_i\,|\,\kappa_{\infty i},\,\gamma_{\infty i})\;,
\end{equation}
and it depends on the galaxy model parameters via the mass model specification 
discussed in Section~\ref{mod-spec}. The logarithm of the likelihood function
is denoted as $l:=\ln{\cal L}$.

\begin{figure}
    \epsfxsize=84mm
    \epsffile{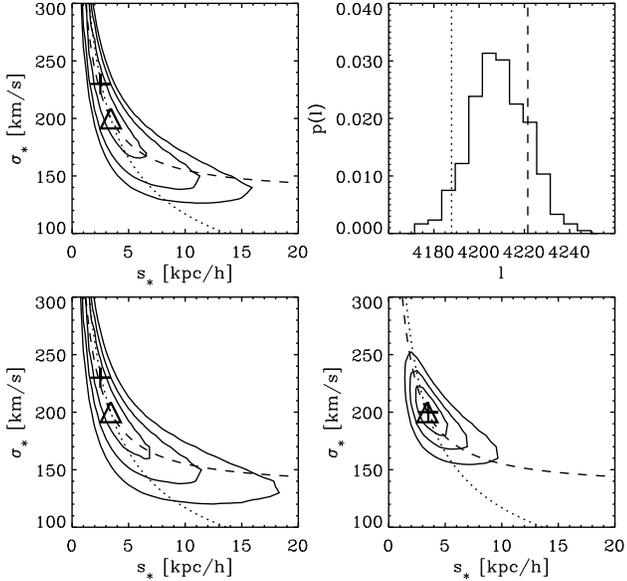}
\caption{Application of the maximum likelihood method to the
simulations with small cut-off radius.
For this particular realization, the analysis includes 3969 background galaxy 
images and 367 cluster galaxies. The {\bf top left} plot displays the 
logarithm of the likelihood as a function of the velocity dispersion 
$\sigma_{\star}$ and the cutoff radius $s_{\star}$. The contours are
$\Delta l:=l-{\rm Max}(l)=-5,\,-3,\,-1$. The triangle denotes the input values 
and the cross marks the maximum of the likelihood function. The dotted line 
connects models with equal total mass, and along the dashed line the mass 
within a projected radius of $5.4h^{-1}\,\rm{kpc}$ is constant. The histogram 
on the {\bf top right} depicts the probability distribution for the value 
of $l$ calculated from the correct mass distribution. The dashed vertical 
line indicates the value ${\rm Max}(l)$ at the maximum of the likelihood
function for the particular realization, and the dotted vertical line
represents the 
likelihood of the reconstruction without galaxies added. For the {\bf bottom} 
diagrams the likelihood contours were transformed into confidence regions as 
explained in the text. The {\bf left} plot is without prior information, and 
the {\bf right} one includes {\it a priori}\/ knowledge of 
$\sigma_{\star}=200\pm15\,\rm{km/s}$. The confidence contours are 99.73\%, 
95.4\%, and 68.3\%.}
\label{like1}
\end{figure}

Fig.\,\ref{like1} demonstrates the application of the likelihood analysis to 
the simulated data for the input model with the small cutoff radius of
$s_{\star}=3.4h^{-1}\,{\rm kpc}$. Here we investigated the dependence of the 
likelihood function on the velocity dispersion $\sigma_{\star}$ and the 
cutoff radius $s_{\star}$, keeping the scaling indices fixed at their input 
values. The analysis includes the background galaxy images from the entire 
field of view, except those which are located very close to cluster galaxies 
and which are therefore likely to be unobservable in practice. More 
specifically, an angular separation limit of $\sqrt{L/L_{\star}}\,3\arcsec$ 
was employed, whereby $L$ denotes the cluster galaxy luminosity. 
Hence, in this study there is no lensing information available on the  
mass distribution of cluster galaxies within a projected radius of 
$D_{\rm d}\,3\arcsec\approx5.4h^{-1}\,{\rm kpc}$. In fact, the likelihood 
contours closely follow the line of models with equal mass within
this radius, and therefore this is the quantity which can be determined best
with this lensing method, while the velocity dispersion and the cutoff radius
cannot be well constrained individually.

In order to verify the absolute likelihood level of the reconstructed mass 
model, the figure also shows the probability distribution for $l$ calculated
with the correct input mass distribution from many different realizations of 
intrinsic background galaxy ellipticities. In accordance with the central 
limit theorem, this histogram is consistent with a Gaussian distribution. 
The maximum of the likelihood function lies well within this distribution, 
and therefore the total mass model, consisting of the reconstructed cluster 
mass distribution plus the galaxy mass model, is statistically consistent with 
the observed image ellipticities.
	
The likelihood contours can be transformed into confidence regions for the
model parameters. The procedure we adopted to achieve this will be
explained at the end of this section. Using only the information provided by 
the lensing analysis allows to set an upper limit on the cutoff radius of 
about $18h^{-1}\,{\rm kpc}$ in the example case depicted in Fig.\,\ref{like1}. 
However, large values for the cutoff radius $s_{\star}$ are only 
compatible with unrealistically low values for the velocity dispersion 
$\sigma_{\star}$. As a consequence, it is possible to achieve much tighter 
limits on $s_{\star}$ by making use of {\it a priori}\/ knowledge on 
$\sigma_{\star}$. If we believe that the measured velocity dispersions of 
elliptical galaxies or the rotational velocities of spirals (divided by a 
factor of $\sqrt{2}$) represent the same quantity as the parameter $\sigma$ 
of the dark matter halo model, we can include this knowledge into the 
analysis. The likelihood function can be regarded as the joint probability 
distribution $p(\epsilon_1,\dots,\epsilon_{N}\,|\,\sigma_{\star},\,s_{\star})$
of observing the image ellipticities for a given set of model parameters. 
According to Bayes' Theorem, the probability distribution for the parameters 
is then given by 
\begin{equation}
p(\sigma_{\star},\,s_{\star})\propto p_{\rm prior}(\sigma_{\star},\,
s_{\star})\,p(\epsilon_1,\dots,\epsilon_{N}\,|\,\sigma_{\star},\,s_{\star})\;,
\end{equation}
and the constant of proportionality is fixed by requiring the proper 
normalization. Confidence regions for the parameters can be found by 
determining the contour lines which enclose a given fraction of the total 
probability. The figure displays the result after taking into 
account the prior information of $\sigma_{\star}=200\pm15\,{\rm km/s}$ which 
makes it possible to derive very interesting limits on $s_{\star}$. 
By specifying a constant prior, the Bayesian reasoning also allows us to 
transform the likelihood contours into confidence regions without 
{\it a priori}\/ information. The corresponding plot in the figure was already
mentioned above. (Strictly speaking, the prior used in this case is constant 
only over the region covered by the plot, but zero for parameter values not
represented by it.)

\subsection{Velocity Dispersion and Cutoff Radius}\label{vel-cut}
\begin{figure*}
    \epsfxsize=84mm
    \epsffile{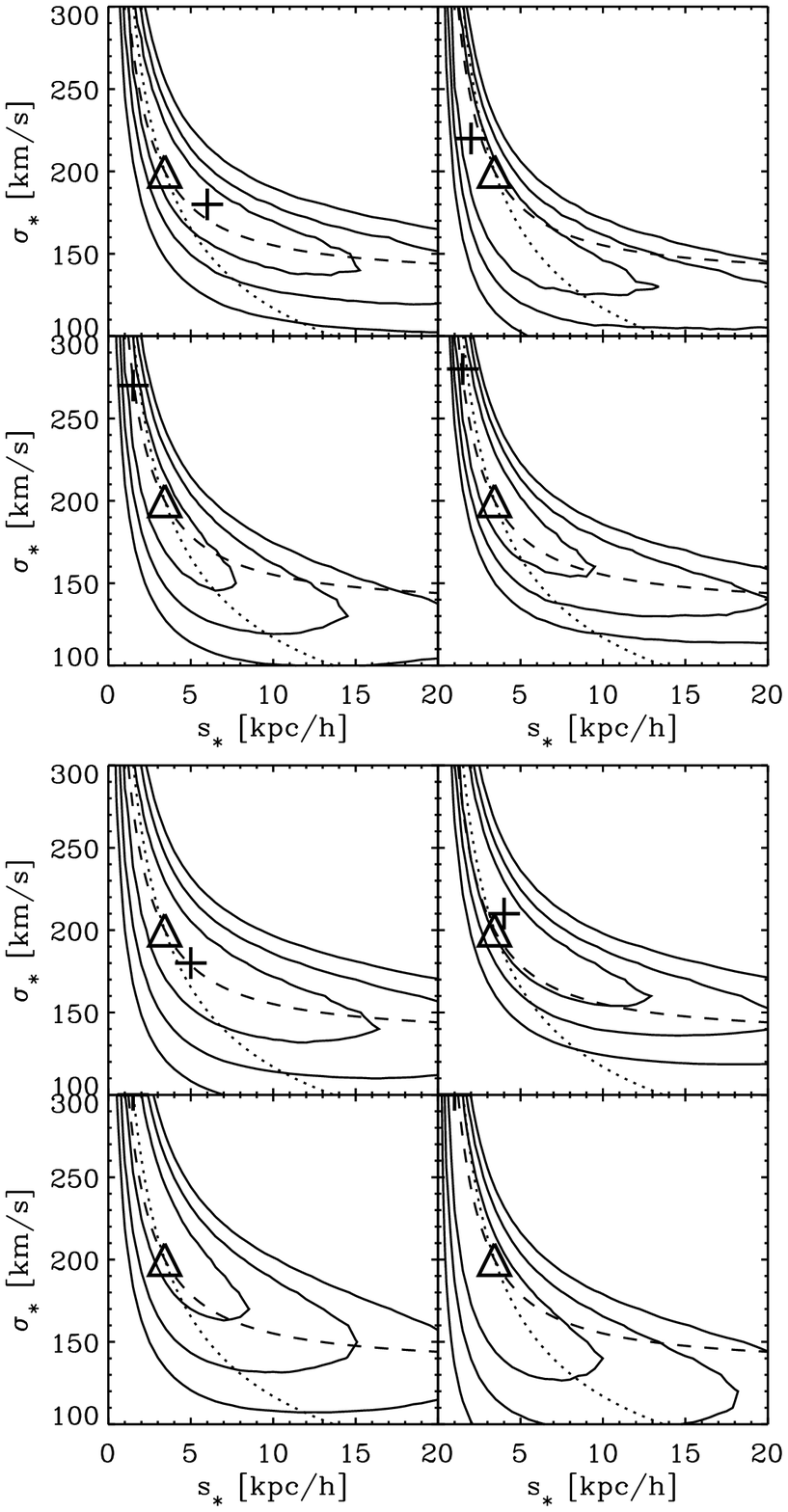}
    \epsfxsize=84mm
    \epsffile{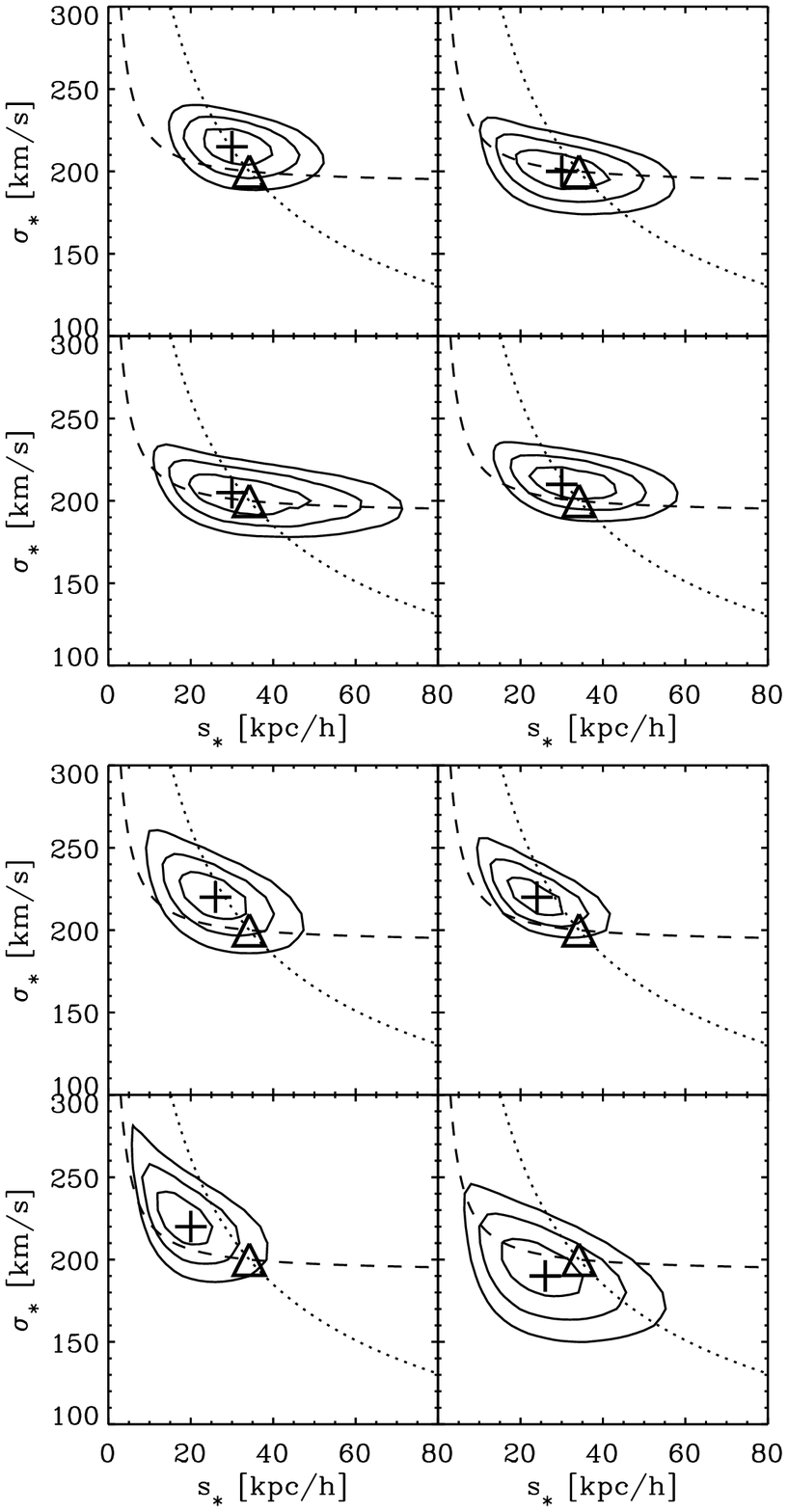}
\caption{Confidence regions for the velocity dispersion and the cutoff radius
for different random realizations. The sets of plots on the {\bf left} are for 
the galaxy input model with a cutoff radius of 
$s_{\star}=3.4h^{-1}\,{\rm kpc}$, and those on the {\bf right} for
$s_{\star}=34h^{-1}\,{\rm kpc}$. Note their difference in the range of
parameter values on the $s_{\star}$-axis. 
As specified in the text, the {\bf top} plots display the results including 
information provided by background galaxy images located in the outskirts of 
the cluster and the {\bf bottom} plots are for the central region. 
The confidence contours are 99.73\%, 95.4\%, and 68.3\%,
determined in the way explained in Section~\ref{like-conf} (without prior 
information), and the meaning of the lines and symbols is the same as in 
Fig.\,\ref{like1}. The random realizations of cluster and background galaxies
are the same as those used for Fig.\,\ref{zeta2}.}
\label{like2}
\end{figure*}

Fig.\,\ref{like2} shows confidence regions (without including prior
information) for the velocity dispersion and the cutoff radius, computed for 
several realizations of cluster and background galaxies.
Here we divided the data into two independent subsets according to the position
of the background galaxy images. In one case we included all images
($\approx3740$) located outside of a square with side length $2\farcm5$ 
centred on the peak of the cluster mass distribution, and the second case 
includes all images ($\approx240$) within this central region. The number of 
cluster galaxies which are located in these areas are $\approx270$ and 
$\approx90$, respectively. Again, the exact numbers are varying for different 
random realizations. 

We start the discussion with the results for the galaxy input model with small 
cutoff radius. In this case, the galaxies contribute only about three per cent 
to the total mass of the galaxy cluster and the mass compensation procedure
mentioned in Section~\ref{mod-spec} is not very important. The figure shows 
that the far fewer images in the centre provide almost the same amount of 
information as the numerous images in the outskirts of the cluster. The 
reasons for this are the higher cluster galaxy density in the centre and the 
significant enhancement of the distortion effects of individual cluster 
galaxies due to the underlying cluster mass distribution. However, the
confidence regions for this central subset of the data depend somewhat
on the details of the reconstruction procedure. In particular, the 
smoothing length must not be too large and the reconstruction grid must not 
be too coarse in order to prevent the cluster's central peak from being 
smeared out in the mass reconstruction.  

For the model with an extended dark matter halo with cutoff radius
$s_{\star}=34h^{-1}\,{\rm kpc}$, about one third of the total mass of the 
system is contained in the galaxies, and therefore the prescription for the 
mass compensation becomes extremely important. In regions of low surface mass 
density, it turned out that the most effective procedure is to subtract mass 
locally at the position of the cluster galaxies in the form of Gaussian 
distributions with widths corresponding to the (local) smoothing scale 
used for calculating the gridded image ellipticities. This is applicable as 
long as the subtracted mass fraction is not exceedingly large. For the subset 
of the data regarding the outskirts of the cluster, the results presented in 
Fig.\,\ref{like2} indicate that the velocity dispersion $\sigma_{\star}$ 
can be retrieved reasonably well in this case, whereas it is more difficult to 
constrain the radial extent of the galaxy mass distribution. 
However, the rather arbitrary mass subtraction procedure is certainly not 
optimal, and the additional uncertainties arising from it are not included 
in the confidence regions, because the maximum likelihood method described 
here assumes that the description of the `global cluster mass distribution'
constructed in that way is correct. Nevertheless, a robust lower limit of 
typically about $10h^{-1}\,\rm{kpc}$ can be set for the cutoff radius, and 
so this model can be distinguished with high significance from the 
low-$s_{\star}$ model discussed above. The upper limits, on the other hand,
are more sensitive to the details of the mass compensation procedure.
(Formally, a strict upper limit can be derived from the fact that the total 
mass is fixed by the cluster reconstruction. In our simulations, for example, 
a galaxy model with $\sigma_{\star}=200\,{\rm km/s}$ and 
$s_{\star}\approx100h^{-1}\,{\rm kpc}$ accounts for all the mass in the 
system.)

For the input model with massive galaxies, it is worth to note that there 
is a very large difference ($\Delta l\approx150$) in the likelihood values 
between the best-fit model with galaxies and the pure reconstruction map 
without any cluster galaxies included, and in contrast to the model without 
extended galaxy halo (see the plot in Fig.\,\ref{like1}) the pure 
reconstruction is not consistent with the observed image ellipticities in an 
absolute likelihood sense. 

The problems of the method become apparent in the diagrams of 
Fig.\,\ref{like2} depicting the results for the central cluster region. 
Here, the input values for the galaxy model parameters cannot be reliably 
recovered, and in addition, the confidence regions change considerably 
when the strategy for the mass compensation is modified. (For the plots shown 
in the figure, the cluster mass reconstruction was scaled down by the mass
fraction put into galaxies in order to conserve the total mass.)
In the outer regions of the cluster, the problem is less severe, because the 
requirements for the accuracy of the cluster mass reconstruction are less 
stringent when the surface mass density is low, and so the method works there 
reasonably well even when the galaxies are massive. In the highly non-linear 
lensing regime of the cluster centre, however, an accurate description of the
cluster mass distribution is essential to obtain reliable results. From our 
investigations we conclude that the cluster reconstruction and the maximum 
likelihood analysis for inferring the properties of the cluster galaxies 
cannot be performed independently, taking the results of the former as an 
input for the latter. Instead, both procedures have to be performed at the 
same time. To this end we employed a maximum likelihood reconstruction of 
the cluster mass distribution in the fashion of Bartelmann et~al. (1996). 
In such a method the presence of cluster galaxies can be taken into account 
explicitly during the reconstruction process. For each set of parameters of 
the galaxy mass model, one can then determine the best representation of the 
underlying cluster mass distribution. Therefore, this approach is also more 
satisfactory in a full maximum likelihood sense. A regularization of the 
cluster mass distribution (for example by an entropy-like term) prevents it 
from exhibiting structures on galaxy scales and thus allows the separation 
between galaxy mass components and the underlying cluster mass 
distribution. That method and its application to our simulations will 
eventually be presented in a separate publication. 

Distinguishing between the dark matter associated with galaxy haloes and dark
matter belonging to a `global cluster mass distribution' poses not only a 
technical problem, but also a conceptual one. Especially towards the centre 
of galaxy clusters where the physical distances between the galaxies become 
very small, making this distinction becomes somewhat artificial, and clearly 
the giant cD-galaxies residing in the centre of many clusters cannot be 
treated with the same formalism as ordinary cluster galaxies.   

Although the general formulation of our method in principle allows the 
treatment of critical clusters as well, we restricted the application in 
this paper to the non-critical case, because otherwise the problems for the 
cluster centre alluded to above would be substantially more serious. For the 
critical regions it is indispensable to model the mass distribution of a 
global component and those of individual cluster galaxies 
simultaneously.\footnote{In the likelihood method of Natarajan and Kneib 
(1997) a description of the `cluster mass distribution' was assumed to 
be available {\it a priori}\/ from the modelling of strong lensing features. 
In their simulations the mass models for the cluster galaxies were added to 
the same known cluster mass distribution for generating the data as well as 
for the analysis.}
In addition, an efficient method should at the same time take into account 
the constraints offered by arcs and multiple image systems, which are then 
likely to be present, as well as the weak lensing information. Again, this 
could be achieved by a maximum likelihood mass reconstruction. 

It was noted for example by Kassiola, Kovner \&~Fort (1993), Wallington, 
Kochanek \&~Koo (1995), Colley, Tyson \&~Turner (1996) or Kneib et~al. (1996), 
that it is necessary to include the effects of individual cluster galaxies in 
order to explain the details of the strong lensing features. By exploiting the 
morphology of strongly distorted images it might well be possible to set 
constraints on the mass distribution of individual cluster galaxies which are 
located close to critical lines. First steps in this direction were made
by Kassiola et~al. (1993) and Wallington et~al. (1995) for two bright
cluster galaxies perturbing the arc system in the cluster 0024+1654.

\subsection{Scaling Parameters}

For calculating the confidence regions shown in Figs.~\ref{like1}
and~\ref{like2} we fixed the scaling parameters for the galaxy haloes at 
their input values ($\eta=4$, $\nu=0.5$). A comprehensive analysis 
within the mass model specified for this method should in principle include 
a maximization of the likelihood over these {\it a priori}\/ unknown 
parameters. This leads to a slight widening of the confidence regions, but 
it does not change any of the general conclusions drawn above.

\begin{figure}
    \epsfxsize=84mm
    \epsffile{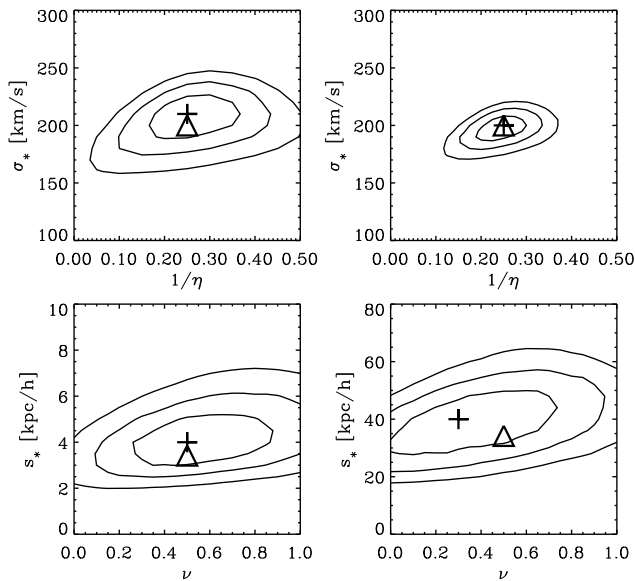}
\caption{Confidence regions for the scaling parameters $\eta$ and $\nu$.
The {\bf left} plots are for the galaxy input model with 
$s_{\star}=3.4h^{-1}\,{\rm kpc}$, taking into account the
information from the total field of view. The {\bf right} plots are for 
the extended halo model ($s_{\star}=34h^{-1}\,{\rm kpc}$) and only 
include the information provided by the background galaxy images located in 
the outskirts of the cluster as specified earlier in the text. 
The significance of the contours and the meaning of the symbols is analogous 
to the previous figures, and the random realization of cluster and background 
galaxies is the same as for the top right plot in each of the panels of 
Fig.~\ref{like2}.}
\label{scalf}
\end{figure}

Here the prospects for determining these scaling indices from the lensing
analysis are briefly mentioned. Fig.~\ref{scalf} displays the confidence 
contours as a function of $\eta$ and $\sigma_{\star}$, and $\nu$ and 
$s_{\star}$. Each time the two remaining parameters were fixed at their 
input values. The plots reveal that the constraints on the velocity dispersion 
scaling index $\eta$ are not particularly tight if the cutoff radius is small.
In order to improve them it would be necessary to add the information from 
several galaxy clusters. In the case of extended dark matter haloes, however,
the information available from one cluster is already sufficient to
derive interesting limits  
on $\eta$, although we excluded the central data region of our simulations 
in this case in order to avoid the problems discussed in the previous section.
Whereas the scaling prescription for the velocity dispersion can be motivated 
by observations as well as by physical arguments, the scaling law for the 
cutoff radius adopted for our analysis is rather hypothetical. In realistic 
applications within galaxy clusters one would expect a much stronger 
dependence of the galaxy halo extent on external effects rather than on the 
luminosity of the cluster galaxy, but apart from binning the data into subsets
we did not investigate more quantitative methods for establishing such 
dependencies. The results on the scaling parameter $\nu$ shown in the figure
should therefore merely be regarded as an indication of the amount of 
information available.

\subsection{Potential Problems}\label{problems}

\begin{figure}
    \epsfxsize=84mm
    \epsffile{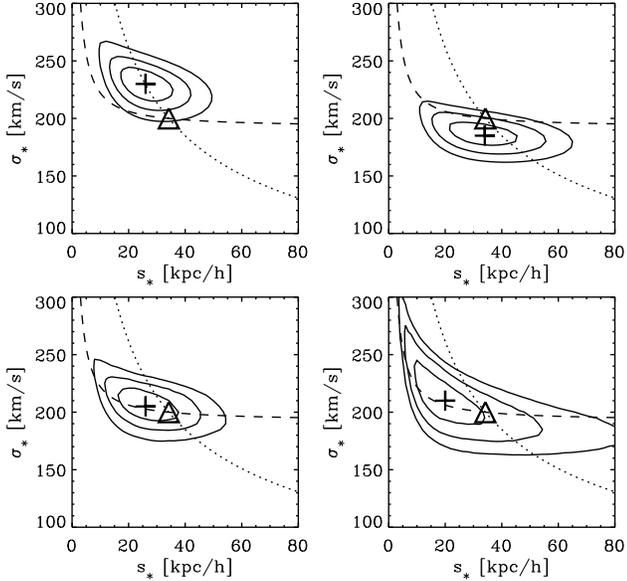}
\caption{The consequences of potential problems. These diagrams should be 
compared with the top right plot in the top right panel of Fig.~\ref{like2}. 
The {\bf top} plots demonstrate the effects of assuming an incorrect redshift
distribution for the source galaxies. The {\bf left} plot is for a distribution
according to equation (\ref{reddist}) with $\zave=0.5$ and $\beta=1$, and the 
{\bf right} one for $\zave=2$ and $\beta=1$.
In the {\bf bottom} plots the analysis only includes cluster galaxies brighter
than $0.3\,L_{\star}$ ({\bf left}), or brighter than $L_{\star}$ ({\bf right}).
The significance of the contours and the meaning of the symbols is
analogous to the previous  
figures.}
\label{probl}
\end{figure}

A general problem for weak lensing studies is the degeneracy between
the distance of the lensed source galaxies and the physical surface
mass density -- and hence the mass -- of the lens. The quantity which
is of importance in this context is the average $\wave$ of the
relative lensing strength, which depends on the redshift distribution
of the source galaxies (for a given cluster redshift and
cosmology). In order to demonstrate the implications of incorrect
assumptions on $p_z(z)$, we repeated the analysis for one of the
diagrams shown earlier. The confidence regions displayed in
Fig.~\ref{probl} were calculated for redshift distributions with
$\zave=0.5$ and $\wave=0.55$, and $\zave=2$ and $\wave=0.84$,
respectively, whereas the true distribution used for generating the
data has $\zave=1$ and $\wave=0.75$. The results confirm that an
underestimate of $\wave$ leads to an overestimate of the velocity
dispersion, and overestimating $\wave$ causes a displacement of the
confidence contours towards smaller velocity dispersions. Note also
that for the low cluster redshift of $z_{\rm d}=0.16$, large
overestimates of the average source redshift $\zave$ do not strongly
affect the results, because the relative lensing strength reaches an
asymptotic value for increasing source redshift. The problem is more
severe for higher-redshift clusters, and in principle specifying the
parameter $\sigma_{\star}$ by other means allows to constrain the
quantity $\wave$. (This is true if the confidence contours are not
`intrinsically' extended in the $\sigma_{\star}$-direction as it is
the case for the low cutoff radius model.) As long as the lens is
non-critical for all redshifts, the results do not strongly depend on
higher moments of the redshift distribution, and any reasonably smooth
function for $p_z(z)$ should be adequate to approximate the correct
description of the probability density distributions for image
ellipticities (see Fig.~\ref{pddeps}). In the critical case, however,
the image distortions carry a large amount of information on the
redshift distribution of the sources, and then the sensitivity of the
results to the choice of $p_z(z)$ has to be taken more seriously.

The additional degeneracy expressed by the mass sheet transformation 
mentioned in Section~\ref{reconstruction} is present even if the average 
lensing strength is known. Although this degeneracy can be broken by other 
means, we would like to remark here that a transformation of the total mass 
distribution (including the galaxies) according to equation (\ref{mass-sheet}) 
merely implies a rescaling of the galaxy mass distributions. Therefore this 
leads to a shift of the confidence regions along the velocity dispersion 
coordinate, and it does not affect the conclusions on the radial extent of the
dark matter haloes.

A potentially serious observational problem for the study described in this
paper could be the reliable identification of cluster galaxies. We tested the 
importance of this issue by ignoring faint cluster galaxies during the 
likelihood analysis. For the respective diagrams of Fig.~\ref{probl} we took
into account cluster galaxies brighter than $0.3\,L_{\star}$ (129 galaxies) or 
brighter than $L_{\star}$ (34 galaxies), whereas the calculation of the image 
shapes of background galaxies included 259 cluster galaxies brighter than
$0.1\,L_{\star}$ within the analysis region for this particular realization.
The results demonstrate that despite their large numbers, the contribution of 
very faint cluster members to the lensing signal is only marginal. The
information contributed by cluster galaxies between $0.3\,L_{\star}$ and
$L_{\star}$, and those brighter than $L_{\star}$ is comparable, because the
smaller numbers of bright and massive galaxies are compensated by their
stronger distortion effects. Neglecting the presence of fainter cluster 
members only increases the noise level and does not systematically affect the 
results. This also indicates that possible small scale clumps in the dark
matter distribution which are not associated to luminous galaxies 
do not bias the results of the likelihood analysis.

Finally, in applications to real observations a distinction could be made 
between spiral and elliptical cluster galaxies because they require different 
normalizations for the velocity dispersion parameter. This can be done 
iteratively, searching for the best solution for one kind of galaxies at a 
time, and should not cause additional problems. However, due to the dominance 
of early-type galaxies, this distinction may be unnecessary in many clusters.

\section{discussion}\label{discuss}

We investigated methods to constrain the mass distribution of cluster 
galaxies from the distortions of the images of faint background
galaxies. In this paper we restricted the treatment to non-critical clusters
(or the non-critical regions of critical ones), and we did not discuss the
observational difficulties in measuring image ellipticities or identifying
cluster galaxies. 

The $\zeta$-statistic is a straightforward method for determining aperture 
masses. Significant (aperture) mass estimates for an ensemble of cluster 
galaxies can be obtained by adding the results for a large number of galaxies.
The method provides a direct handle on the lensing signal of the cluster
galaxies without the need to specify a model for their mass distribution.
The galaxy lensing effects are amplified by an underlying cluster mass 
distribution. Hence, in regions with non-negligible surface mass density, a 
cluster mass reconstruction is necessary in order to take this effect into 
account in the calculation of the $\zeta$-statistic. Due to geometrical 
limitations it is not possible to include all available information into the 
method. Towards the cluster centre the increasing number density of cluster 
galaxies precludes a useful application of the $\zeta$-statistic. In addition, 
the generalization into the non-linear regime also implies an increasing 
sensitivity to uncertainties in the description of the cluster mass 
distribution or the redshift distribution of the source galaxies. In the 
outskirts of clusters, however, the $\zeta$-statistic is applicable without 
major technical difficulties.

For a quantitative analysis, a maximum likelihood method is more appropriate.
We tried to separate the treatment of cluster galaxies and a `global cluster 
mass distribution' by reconstructing the latter one using standard inversion 
methods and then adding parametrized mass models for the galaxies on top of 
that. The results of our simulations demonstrate that this method is reliable 
-- in the sense of correctly retrieving the input parameters for the 
galaxy mass models within their confidence regions -- as long as the mass in 
galaxies is small compared to the total mass of the system. However, if the 
cluster galaxies do have extended dark matter haloes, this is not the case. 
The potentially significant mass fraction contributed by them also shows up in 
the cluster mass reconstruction, and adding additional mass in the form of
galaxy models would violate the total mass constraint given by the 
reconstruction. We dealt with that problem by applying empirical and admittedly
inelegant mass compensation procedures. This approach turned out to be 
workable, though not completely satisfying, in the outskirts of clusters 
where the requirements on the accuracy of the description of the cluster mass 
component are moderate. 

In the highly non-linear region of the cluster centre, however, it is 
impossible to treat the image distortion effects caused by a global mass 
component and those caused by individual cluster galaxies independently. 
Rather, the principle of maximum likelihood should be taken seriously and the 
method of choice should allow to determine the best description of the cluster 
mass component for each given set of galaxy model parameters by explicitly 
taking the presence of the galaxies into account. In general, this cannot be 
accomplished by resorting to simple parametrized mass models for the cluster 
component itself. These represent an unjustified restriction and could 
therefore severely bias the results. Observational as well as numerical work 
indicates that clusters of galaxies cannot be regarded as nicely virialized 
systems. Instead, their mass distribution often exhibits complicated 
morphologies and hence a virtually parameter-free approach is warranted for 
describing them. We developed a generalized maximum likelihood method which
enables us to cope with the problems discussed above and we will report on 
our experience therewith elsewhere.

In those cases for which we classified the likelihood method presented in 
this paper as reliable or applicable, we believe that the general picture 
provided by the confidence regions in the galaxy model parameter space is 
correct. Nevertheless, introducing additional degrees of freedom by allowing 
the cluster component to adapt to changes of the galaxy model will 
tend to widen the confidence regions. Especially for the galaxy input model 
with extended dark matter haloes an even more pronounced elongation of the
confidence regions along the cutoff radius coordinate in the model parameter
space can be expected. In a reliable analysis for extended galaxy haloes in 
the central cluster region, this should then be the case as well. This 
model degeneracy between mass in galaxies and mass in a global cluster 
component also reflects the conceptual difficulties of making a clear-cut 
distinction between the two. Without further information on the dynamical 
state of this matter, it might be more appropriate to interpret the results 
on the cutoff radius of galaxies near the cluster centre as a characteristic 
scale of mass clumping around cluster galaxies rather than necessarily as 
`the extent of the galaxy mass distribution'. 

In our simulations we specified two extreme models for the cutoff
radius of cluster galaxies, and in the discussion we concentrated on
investigating the capabilities of reliably retrieving these
well-defined input values for the mass model parameters. Although it
is difficult to tightly constrain `the extent of the dark matter
haloes' from the information available, it is nevertheless feasible to
distinguish with high significance between extreme models with or
without extended dark matter halo, as well as to detect possible
spatial variations of the galaxy halo properties. In realistic
situations, a dependence on the physical distance from the cluster
centre or the (three-dimensional) density of the environment could be
expected, whereas the observations only allow direct access to the
projected distance and the surface mass density. For example, if the
extent of the galaxy halo is determined by the density of the environment, a
linear dependence of the cutoff radius on the physical distance from the
cluster centre could be expected for clusters with isothermal mass profile.
In more sophisticated simulations a dependence of the cutoff radius as a 
function of the density of the environment or the distance from the cluster 
centre could be explicitly included. This would allow to develop effective
strategies for quantifying possible trends of the characteristic
extent of galaxy haloes and to assess the uncertainties introduced by
projection effects. For a more direct comparison of observable effects
with theoretical predictions, another option is to use high resolution
N-body simulations in which cluster galaxies can be resolved
individually.

Although the analysis in this paper was confined to a single cluster, the 
results from different clusters can easily be combined statistically.
The observational prospects of weak lensing studies have been widely
discussed in recent years, and several observations which are suitable for 
carrying out the kind of project described here are already
available. 

Shortly before the submission of this paper, a preprint was put on 
the web by Natarajan et~al. (1997). They applied the methods of Natarajan 
\&~Kneib (1997) to HST images of the cluster AC114, detected a 
galaxy-galaxy lensing signal, and obtained an estimate of 
$\approx 10h^{-1}\,{\rm kpc}$ for the size of an $L_*$-galaxy halo. 
When compared to the results of Brainerd et al. (1996) this provides 
an indication that the haloes of cluster galaxies are less extended than 
those of field galaxies.

\section*{acknowledgments}
We thank Matthias Bartelmann for making the N-body cluster simulation 
available, Stella Seitz for discussions, and Simon White for comments on 
the manuscript. This work was supported in part by the Deutscher 
Akademischer Austauschdienst (Doktorandenstipendium HSP III) and the 
Sonderforschungsbereich 375-95 der Deutschen Forschungsgemeinschaft.

\label{lastpage}

\end{document}